\documentclass[letterpaper, 10 pt, conference]{ieeeconf}  
\usepackage{comment}
\usepackage{cite} 
\let\labelindent\relax
\usepackage{enumitem}

\usepackage{amssymb, mathbbol, mathtools, amsthm} 
\usepackage{graphicx}
\usepackage{csquotes} 
\usepackage{hyperref} 
\usepackage{xcolor} 
\usepackage{textcomp}
\usepackage[export]{adjustbox}
\usepackage{subcaption}
\usepackage{setspace}
\usepackage{float}
\usepackage{array}
\usepackage{tikz}
\usepackage{graphics} 
\usepackage{epsfig} 
\usepackage{caption}
\usepackage{mathptmx} 
\usepackage{times} 
\IEEEoverridecommandlockouts                              
\newtheorem{theorem}{Theorem}
\newtheorem{lemma}{Lemma}
\theoremstyle{definition}
\newtheorem{definition}{Definition}
\newtheorem{example}{Example}
\theoremstyle{remark}
\newtheorem{remark}{Remark}
\theoremstyle{corollary}
\newtheorem{corollary}{Corollary}
\title{\LARGE \bf {On Real Eventual Exponential Positive property of complex-valued Laplacians and their pseudoinverses} }
\author{Aditi Saxena$^{1}$, Twinkle Tripathy$^{1}$ and Rajasekhar Anguluri$^{2}$
\thanks{$^{1}$Aditi Saxena is a research scholar and Twinkle Tripathy is an assistant professor at the Department of Electrical Engineering, Indian Institute of Technology Kanpur, India. (email: {\tt\small saditi23@iitk.ac.in, ttripathy@iitk.ac.in)}}%
\thanks{$^{2}$Rajasekhar Anguluri is an assistant professor at the Centre for Systems and Control,  Indian Institute of Technology Bombay, India. (e-mail: {\tt\small rajanguluri@iitb.ac.in}).}}
\begin{document}
\maketitle
\thispagestyle{empty}
\pagestyle{empty}
\begin{abstract}
The property of \emph{real eventual exponential positivity} guarantees {\color{black} the state trajectories of a linear time-invariant system} to evolve asymptotically to right half of the complex-plane{\color{black}\cite{saxena2025laplacianflowscomplexvalueddirected}}. \textcolor{black}{It has applications in the characterization of herdability and reachability of states in networked systems. Networks with complex-valued edge-weights arise in AC power systems and opinion dynamics. Recent work has introduced real eventual exponential positive (rEEP) Laplacians, for which the trajectories of the associated network flow eventually converge to consensus. We establish further properties of rEEP Laplacians and their pseudoinverses. Firstly, we show that the rEEP property for complex-valued matrices exhibits invariance under specific rotational transformations. Secondly, we establish that the property of rEEP is preserved under the Moore--Penrose pseudoinverse operation, highlighting its robustness under algebraic manipulations. These results apply to effective impedance computation in AC networks and to consensus in multi-agent systems.} 
Finally, we provide numerical examples to validate our theoretical results. 
\end{abstract}\vspace{-2mm}
\section{INTRODUCTION}

{\color{black}Complex-valued networks, whose edge weights are complex numbers, arise in quantum physics, electromagnetics, power and societal systems\cite{bottcher2024complex}. Although complex-valued Laplacians have appeared in formation control and spectral graph theory \cite{DONG20161,dong2014complex}, dynamics of such networks remain far less understood than their real-valued counterparts \cite{dorfler2012synchronization,olfati2007consensus}. The main obstacle is largely structural. For Hermitian weights, the familiar tools of real symmetric matrices largely survive. For complex symmetric or edge asymmetric weights, which arise in AC power networks and recently in opinion dynamics \cite{saxena2026emotionsspreadaffectdecisionmaking}, these tools no longer apply: eigenvalues are complex, positive semi-definiteness is undefined, and the Perron–Frobenius (PF hereafter) theorem no longer applies. 

Eventual positivity \cite{fontan2022multiagent} has proven useful for signed real networks, where it recovers PF-type conclusions without nonnegative weights \cite{altafini2019investigating, fontan2021properties}. Recent literature has pushed this idea to networks with complex-valued edge weights \cite{lin2013leader, aditiECC2025,tian2026generalizing,saxena2025laplacianflowscomplexvalueddirected}. Building on these PF-type arguments \cite{varga2012}, these papers introduced rEEP Laplacians, for which the real parts of steady-state trajectories eventually become positive. Interestingly, \cite{aditiECC2025} studied the dynamics and applications of the associated pseudoinverse flows, and analyzed them on undirected and weight-balanced digraphs. }

This paper relaxes the assumption of undirected and weight-balanced digraphs and analyze the pseudoinverse Laplacian flows for non-weight-balanced digraphs. Further, our emphasis is on the property of eventual exponential positivity (EEP hereafter). In contrast to a real-valued system, where EEP ensures the state-transition matrix to have all positive entries after a finite time and to remain so thereafter; for a complex-valued system, entrywise positivity is no longer sufficient to characterize rEEP, since each non-zero complex entry possesses both a real and an imaginary part. Thus, the property of rEEP in complex-valued networks guarantees positive real parts in the matrix exponential at the steady state. This property depends on the eventual properties of matrices and is of great interest in control theory in analyzing the state evolution of linear systems. It has significant implications in determining the conditions for herdability and reachability to the $\mathbb{C}_+$ plane (see Sec.\ref{sec:preliminaries}) for networked control systems.

We compute the Moore-Penrose pseudoinverse of the Laplacian matrix to obtain the pseudoinverse Laplacian. Building upon our previous work \cite{saxena2025laplacianflowscomplexvalueddirected}, we extend the notion of EEP from real-valued systems to complex-valued systems. \textcolor{black}{To the best of our knowledge, we are the first to answer the following question for complex-valued networks: When does the trajectory from an initial point $x_0$ become (entrywise) real positive and remain real positive for all time thereafter in a linear system? }Below, we summarize our key contributions:
\begin{enumerate}
\item We establish that the property of rEEP is invariant under certain global rotations of Laplacian matrices under some mathematical conditions. Equivalently, if all the complex-valued edge-weights are rotated by a given angle, then the resulting Laplacian matrix still satisfies the property of rEEP.  Further, we show that rEEP can be characterized through real dominant eigenvectors only.
\item  We derive the necessary and sufficient conditions to prove that the property of rEEP is closed under the operation of pseudoinverse in complex-valued digraphs. We use complex PF theory to establish the conditions for the rEEP property of the matrices. 
\end{enumerate}
 
 \textit{Related Literature:} In \cite{altafini2019investigating},\cite{fontan2021properties}, the authors show the inter-relation between the property of EEP and the spectrum of Laplacian matrices for signed networks with real edge-weights. Recently, researchers have started exploring complex-valued Laplacians; the authors provide necessary and sufficient conditions for the Laplacian matrix to be singular, based on the topology of graphs in \cite{DONG20161}. Further, we introduce the rEEP property of complex-valued Laplacians in \cite{aditiICC2024} and show that this property ensures consensus in complex-valued digraphs \cite{saxena2025laplacianflowscomplexvalueddirected}. The pseudoinverse Laplacian flow systems for complex-valued networks are introduced in \cite{aditiECC2025} for undirected and weight-balanced digraphs. Moreover, the notions of structural balance and anti-balance for Hermitian matrices are proposed in \cite{tian2024structuralbalance}. 

\textcolor{black}{Our existing works introduced the rEEP property and established its role in studying consensus in Laplacian flows. In contrast, the present work investigates the implications of rEEP property in the pseudoinverse Laplacian flows beyond the class of undirected and weight-balanced digraphs. Additionally, the invariance of this property under rotations of complex-valued Laplacian matrix and the possible rotations in eigenvectors are discussed here; unlike the previous works}.
\section{PRELIMINARIES}\label{sec:preliminaries}
Let $\mathbb{C}^{n \times n}$ denote the space of $n\times n$ complex-valued matrices. For $n=1$, the space is $\mathbb{C}$. Let ${\Re}(m) \text{ and } {\Im}(m)$ denote the real and imaginary parts of $m \in \mathbb{C}$, respectively. The imaginary unit $\iota=\sqrt{-1}$. The right half of the complex- and real-planes  are $\mathbb{C}_{+} = \{ {\Re}(m) \geq 0, \text{ for all } m \in \mathbb{C}\}$ and $\mathbb{R}_{+} = \{ m \geq 0, \text{ for all } m \in \mathbb{R}\}$, respectively, where the open right half of complex plane is denoted as ORHP. $\mathbb{1}_n$ and $\mathbb{0}_n$ are all ones and zeros vectors. 
\subsection{Matrix Algebra}
 A matrix $M \in \mathbb{C}^{n \times n}$  is real positive if $\Re(m_{ij}) > 0$ for all the entries 
(hereafter $\Re(M) > 0$). A matrix is real eventually positive (rEP hereafter) if $\Re(M^{k}) > 0 $ for some $ k\geq k_{0}$. 
The complex conjugate transpose and transpose of matrix $M$ are $M^{H}$ and $M^{T}$, resp. The pseudoinverse of $M$ is denoted by $M^{\dagger}$. 
The spectrum of $M$ is $\operatorname{spec}(M)=\left \{ \lambda_{1},\lambda_{2},...,\lambda_{n} \right \}, \text{ where } \lambda_{i}$ is the $i^{th}$ eigenvalue of $M$. An eigenvalue having the largest magnitude is the dominant eigenvalue and is denoted by $\lambda_{1}$. The corresponding eigenvectors are referred as dominant eigenvectors \cite{varga2012}. The right and left dominant eigenvectors are denoted by $v$ and $w^H$, respectively. The possible angle of rotation for the dominant eigenvector is represented as $\psi$. 

Now, we cull some important results, which play a crucial role in our analysis. More details are provided in \cite{varga2012,saxena2025laplacianflowscomplexvalueddirected}. 
\begin{definition}\label{def2:strong PF}(\textbf{\textit{Strong Complex PF Property\cite{varga2012}}}) A matrix $M \in \mathbb{C}^{n\times n}$ has the strong complex Perron-Frobenius property if its dominant eigenvalue $\lambda_{1}$ is positive and simple with $\lambda_{1}>\left|\lambda_{i}\right|$ for all $i \in \left \{2,...,n \right \}$. The right eigenvector ${v}\in \mathbb{C}^n$ associated with $\lambda_1$ is such that  $\Re({v})>\mathbb{0}_n$.
\end{definition}
\begin{definition}\label{def:real EEP}(\textbf{\textit{Real Eventually Exponentially Positive matrix (rEEP)\cite{aditiICC2024}}}) A matrix $M \in \mathbb{C}^{n\times n}$ is rEEP if there exists an exponential index $\tau_{0}$ such that $\Re(e^{M \tau})>0$ for all $\tau \geq \tau_{0}$.
 \end{definition}
 \begin{definition}\label{def:pseudoinverse}
(\textbf{\textit{Moore-Penrose pseudoinverse \cite{2012moore}}})
The Moore-Penrose pseudoinverse of $M \in \mathbb{C}^{n\times m}$ satisfies \cite{2012moore}: 
\begin{enumerate}[label=(\roman*)]
     \item  $M M^{\dagger}M=M$,\label{eq1:Linv}
     \item $M^{\dagger}M M^{\dagger}=M^{\dagger}$, \label{eq2:Linv}
     \item  $M M^{\dagger} \in\mathbb{C}^{m \times m}$ and $M^{\dagger}M \in \mathbb{C}^{n \times n}$ are self-adjoint. \label{eq3:Linv} 
 \end{enumerate}\end{definition}
 \begin{lemma}\label{thm:non wt.balanced}
 (\textbf{\textit{Marginal Stability of $L \in \mathbb{C}^{n \times n}$\cite{saxena2025laplacianflowscomplexvalueddirected}}})
Consider a strongly connected digraph $\mathcal{G}(A)$ and for $1\leq i,j\leq n$. 
Let $v=\alpha \mathbb{1}_n, \alpha \in \mathbb{C} 
\text{ and } w \in \mathbb{C}^n$ be the right and left eigenvectors associated with the zero eigenvalue of $L$, respectively. Suppose the phase angles, $ \phi_{vi} = \angle (v)_i \text{ and } \phi_{wi} = \angle (w)_i $ lie in $[-45^{\circ},45^{\circ}]$. Then, the following statements are equivalent:
  \begin{enumerate}[label=(\roman*)]
      \item $-L$ is rEEP.
      \item $-L$ is marginally stable of corank $1$.
    \end{enumerate}
\end{lemma}
\textcolor{black}{Lemma~\ref{thm:non wt.balanced} relates the rEEP property with the marginal stability of the complex-valued Laplacian matrix. This lemma is instrumental to the analysis of pseudoinverse flows.} 
\subsection{Graph Theory} 
\looseness=-1
Let $\mathcal{G}(A)=(\mathcal{V},\mathcal{E},A)$, where $\mathcal{V}$ be the network or graph of interest and $\mathcal{E}$ are the sets of vertices and edges respectively. The phase angles $\theta_{ij}\in \mathbb{C} \text{ and } \phi_k \in \mathbb{C}$ are associated to the directed edges from nodes `$i$' to `$j$' and `$k^{\text{th}}$' components of the eigenvector. The adjacency matrix is $A \in  \mathbb{C}^{n \times n}$. 
In a strongly connected directed graph (digraph) there exists a directed path between any pair of vertices. In digraphs, out-degree of a vertex is the sum of edge-weights of all the outgoing edges, whereas in-degree is the sum of edge-weights of incoming edges. Collectively for all nodes, the degree matrices are given by $D_{out}=\operatorname{diag}(A\mathbb{1}_{n})$ and $D_{in}=\operatorname{diag}(A^{T}\mathbb{1}_{n})$\cite{bullo2018lectures}. 
The Laplacian matrix is defined as $L=D_{out}-A$. 

{\color{black} We make use of polar as well as rectangular form representation of the complex number. Thus for $A \in \mathbb{C}^{n\times n}$, with non-zero $a_{ij}$, we let 
$a_{ij}=r_{ij} \angle \theta_{ij}$, where $r>0$ is the magnitude and $\theta_{ij} \in [0^{\circ},360^{\circ})$ ({\color{black} measured in degrees}) is the phase angle. The angles of edge-weights may represent social interactions among agents \cite{saxena2026emotionsspreadaffectdecisionmaking}.} 
\subsection{{Problem Formulation}}
Consider a $\mathcal{G}(A)$ with $n$ nodes with complex-valued edge-weights.
 Let $x(\tau) \in \mathbb{C}^n$ be the state vector and consider the (generalized) Laplacian flow system
\begin{equation}\label{eq:Lflows}
    \dot{x}(\tau)=-\bar{L}x(\tau), \quad \tau \geq0, 
\end{equation}
where $\bar{L}:\mathbb{C}^{n\times n}\to \mathbb{C}^{n\times n}$. We consider the cases where $\bar{L}=L$ and $\bar{L}=L^{\dagger}$. 

We are interested in studying the relation between the spectrum of the Laplacians and the pseudoinverse Laplacians. \textcolor{black}{Specifically, we establish that rEEP Laplacians imply rEEP pseudoinverse Laplacians. Also, this property remains unchanged when all the edge-weights are rotated by an angle under some mathematical conditions. These results hold for a class of complex-valued networks, which are characterized by real dominant eigenvectors. }Real dominant complex-valued networks arise in settings like phase-dependent interactions, distributed coordination, power system synchronization, and wireless communication. We use complex PF theory (see Def.\ref{def2:strong PF}) to derive our results. 
\section{Main Results}\label{sec:complex matrix}
We begin by establishing relationships between the Perron-Frobenius property in Def.\ref{def2:strong PF} and the proposed notion of rEEP in Def.\ref{def:real EEP}. Our approach relies on the spectral decomposition of complex matrices and is inspired by \cite{varga2012}. These relationships serve as a starting point for analyzing reachability in complex-valued networks. The rEEP matrices have significant applications in control theory to study reachability\footnotemark[1] and holdability\footnotemark[2] of $\mathbb{C}_{+}^{n}$. 
 \footnotetext[1]{The reachability of $\mathbb{C}_{+}^{n}$ refers to the state trajectory becoming positive in a finite time.} 
\footnotetext[2]{The holdability of $\mathbb{C}_{+}^{n}$ refers to the state trajectory becoming positive and remaining so for all subsequent times.} 

Let the eigenvalues of $M \in\mathbb{C}^{n\times n}$ be arranged as $|\lambda_1|\geq \ldots \geq |\lambda_n|$ and consider the set of matrices:
\begin{align}
\label{eq:strongPFmatrix}
\hspace{-2mm} \mathcal{PF} &\!=\! \{M \in\mathbb{C}^{n\times n}:~ \lambda_1(M) \in \mathbb{R}_{+}\text{ is simple }, \Re{(v)}\!>\!0\}.
\end{align} 
    \textcolor{black}{In the class of matrices belonging to $\mathcal{PF}$, we know that they have a simple dominant eigenvalue $\Re{(v)}>0$. These matrices are relevant in studying the stability of complex-valued linear systems. Specifically, PF matrices are used in power system small-signal stability analysis, consensus, and synchronization in multi-agent networks. However, the PF property of complex-valued matrices alone does not guarantee the steady-state behaviors, because of the complex-valued entries in the dominant eigenvectors. To address this gap, the existing literature has used an additional property of real dominance for dominant eigenvectors  \cite{varga2012,saxena2025laplacianflowscomplexvalueddirected}.}

The \emph{real dominance} property plays a significant role in the analysis of complex-valued matrices. It requires the real part of each dominant eigenvector component to have a greater magnitude than its corresponding imaginary part. The dominant eigenvectors of a complex-valued matrix may or may not satisfy this property. However, owing to the rotations inherent in complex-valued eigenvectors, the components can be suitably rotated to satisfy real dominance. Thus, we consider an angle of rotation `$\psi$' to account for the rotation by which we rotate the components of eigenvector satisfy \emph{real dominance}. Nevertheless, such an angle of rotation does not always exist for certain matrices. We provide an example below where the angle of rotation does not exist, and hence we cannot rotate the left dominant eigenvectors so that all their components lie in the permissible region. 
 \begin{example}\label{ex:tannotsatisfied}
Consider a matrix $M$ where,  
\begin{equation*}
M=\begin{bmatrix}
35-10\iota &0  & 30+10\iota \\
100-20\iota &-35+20\iota & 0\\ 
0 & 1-4\iota & 64+4\iota \\ 
\end{bmatrix}.
\end{equation*} 
The eigenvalues of matrix, $M$ are $\operatorname{spec}(M)=\{ -33.56 +19\iota,32.56-5\iota,65 \}$. The dominant right eigenvector $v$ is real-valued, whereas the normalized left eigenvector is, $w^{H}=[0.019-0.22\iota ~0.04-0.05\iota~ 1.67+0.2\iota]$, 
 and the left eigenvector is $z^{H}=[-0.009+0.12\iota ~0.017+0.036\iota~ 0.99]$. Both the left eigenvectors do not have the property of \emph{real dominance} when we geometrically imagine each element of $w^H \text{ and } z^H$ as a vector in the complex plane. Then, none of these vectors lie in the region $[-45^{\circ},45^{\circ}]$ . Also, there does not exist a possible angle of rotation $\psi$ since $\angle(w^H)=[-85^{\circ}~-55.2^{\circ}~9.4^{\circ}] \text{ and } \angle(z^H)=[129.5^{\circ}~ 64.7^{\circ}~0^{\circ}]$. 
 \end{example}
The eigenvectors of complex matrices allow for a degree of rotation, resulting in a broader set of matrices. However, the existence of a rotation angle that ensures real dominance is not always guaranteed. The following lemma addresses this gap and ensures the existence of an angle of rotation.
\begin{lemma}\label{lemma:existence of rotation}
  Consider a vector `$m$' with its components in polar form. Let $\phi_1 \text{ and } \phi_2$ be the smallest and largest phase angles of all the components. Then, there does not exist an angle `$\psi$' such that $m e^{\iota  \psi}$ lies in the region formed by $ \left [-\frac{\pi}{4} , \frac{\pi}{4} \right ]$ if $|\left (\phi_1- \phi_2 \right )|\geq \frac{\pi}{2}$ radians. 
\end{lemma}
Using Lemma~\ref{lemma:existence of rotation}, \textcolor{black}{we conclude that if the components of dominant eigenvector of lie away from each other, then it is not possible to rotate them so that all the components remain in the desired region. }Consequently, it cannot be guaranteed that the matrix exhibits rEEP. 

The next result discusses the significance of \emph{real dominance} {\color{black} for establishing rEEP matrices}.




\begin{theorem}\label{thm:stronglemma1}
Let \( M \in \mathbb{C}^{n \times n} \). Suppose the normalized dominant right and left eigenvectors \( v \) and \( w \) satisfy \( \Re(v) \geq |\Im(v)| \) and \( \Re(w) \geq |\Im(w)| \). If \( (M + dI) \) and \( (M + dI)^H \in \mathcal{PF} \), then \( \Re((M + dI)^k e^{\iota \theta}) > 0 \) for all \( k \geq k_0 \) for some value of \( \theta \) and \( d \geq 0 \). 
\end{theorem}
\begin{proof}  
\textcolor{black}{We have $(M+dI) \text{ and } (M+dI)^H \in \mathcal{PF}$. 
Consider} the case where the magnitude of real part of all the components of unnormalized left eigenvector \( z \) is greater than or equal to its imaginary part. We consider two cases. 




\emph{\textcolor{black}{Case $(i):$} Analysis under the condition $\Re(z) \geq |\Im(z)|$}

The Jordan decomposition of a matrix is given as $M=V J W^H$, where $J$ is the Jordan matrix consisting of all the eigenvalues. Let $\lambda_1$ be the dominant eigenvalue in the square matrix $J$. 
Then, $M$ is partitioned as:
$$
M=\left[v \mid V_{n, n-1}\right]\left[\begin{array}{l|l}
\lambda_{1} & 0 \\
\hline 0 & J_{n-1}
\end{array}\right]\left[\begin{array}{l}
w^{H} \\
\hline W_{n-1, n}
\end{array}\right],
$$
where $v \text{ and } w$ are the right and left dominant eigenvectors of $M$, normalised such that $w^{H} v=1$. {Since}  $(M+dI) \text{ and } (M+dI)^{H} \in \mathcal{PF}, \text{ this implies } \Re(v)>0 \text{ and }\Re(w)>0$ from Equation~\eqref{eq:strongPFmatrix}. 
As per hypothesis $(M+dI)\text{ and }(M+dI)^{H} \in \mathcal{PF}$,
\begin{equation}\label{eq: jordan i implies ii}
\frac{(M+dI)^{k}}{\left(\lambda_{1}+d\right)^{k}}=\left[v \mid V_{n-1, n}\right]\left[\begin{array}{c|c}
1 & 0 \\
\hline 0 & \frac{(J'_{n-1})^{k}}{\left(\lambda_{1}+d\right)^{k}}
\end{array}\right]\left[\begin{array}{c}
w^{H} \\
{W_{n-1,n}}
\end{array}\right],
\end{equation}
where $J'_{n-1}=\operatorname{diag}(\lambda_2+d,\ldots,\lambda_n+d)$ 
and $\lambda_{i} \in \operatorname{spec}(M)$.

We have $z \text{ such that } w=c z=\alpha  e^{\iota \theta} z $ where $\alpha \in \mathbb{R}_{+}$ as both $w \text{ and } z$ are dominant left eigenvectors of $M$. Then, $z^{H} v=\frac{e^{\iota \theta}}{\alpha}$ has positive real part as $\Re(v) \geq |\Im(v)|$ \text{ and } $\Re(z) \geq |\Im(z)|$ holds with the strong complex PF property. 
For some $\theta \in \mathbb{R}$, the following holds by substituting $w^H=c z^H=\alpha  e^{-\iota \theta} z^H$:
\begin{equation}\label{eq:normalisedvectors}
\lim_{k \rightarrow \infty} \frac{(M+d I)^{k}}{\left(\lambda_{1}+d\right)^{k}}=v w^{H} =\alpha v e^{-\iota \theta} z^{H}, 
\end{equation} 
\begin{equation}\label{eq:psi=0} 
\lim_{k \rightarrow \infty} \frac{(M+d I)^{k}e^{\iota \theta}}{\left(\lambda_{1}+d\right)^{k}}=\alpha vz^{H}. 
\end{equation} 
Thus, due to the assumption on the eigenvector $z$, we conclude $\Re(\alpha vz^{H}) > \mathbb{0}_{n\times n}$ which in turn ensures \( \Re((M + dI)^k e^{\iota \theta}) > 0 \).

\textcolor{black}{\emph{Case $(ii):$} Analysis under the condition: $\Re(z)  \ngeqslant  |\Im(z)|$}

Since the real {parts} of the components of vector $z$ are less than their imaginary counterparts, we choose a real $\psi$ such that \( \Re(ze^{\iota \psi}) \geq |\Im(z e^{\iota \psi})| \) holds. Therefore, $ w=\alpha  e^{\iota (\theta-\psi)} z $ gives $ w^{H}=\alpha  e^{-\iota (\theta-\psi)} z^H$, where the angle $\theta$ is the phase difference between both the left eigenvectors. From Equation~\eqref{eq:normalisedvectors},
\begin{align*}
\lim_{k \rightarrow \infty} \frac{(M+d I)^{k}}{\left(\lambda_{1}+d\right)^{k}}=v w^{H}, 
\end{align*} 
Substituting $ w^{H}=\alpha  e^{-\iota (\theta-\psi)} z^H$, we get
\begin{align*}
\lim_{k \rightarrow \infty} \frac{(M+d I)^{k}}{\left(\lambda_{1}+d\right)^{k}}=v w^{H}= \alpha  e^{-\iota (\theta-\psi)}v z^H ~ \forall~{\psi \in [-\psi_a,\psi_b]},
\end{align*}
where $[-\psi_{a},\psi_{b}]$ contains values of the angles of rotation by which we may rotate the unnormalized left eigenvector such that \( \Re(ze^{\iota \psi}) \geq |\Im(z e^{\iota \psi})| \) follows. 

Finally, 
\begin{equation}\label{eq:rotated z}
\lim_{k \rightarrow \infty} \frac{(M+d I)^{k}e^{\iota (\theta-\psi)}}{\left(\lambda_{1}+d\right)^{k}}=\alpha v z^{H}~ \forall~{\psi \in [-\psi_a,\psi_b]}. 
\end{equation}
From Equation~\eqref{eq:rotated z}, we conclude that the real part of $(M+dI)^k e^{\iota (\theta-\psi)}$ is eventually positive as the magnitude of real part of rotated left eigenvector $z^H$ is greater than the magnitude of imaginary part in the eigenvectors. 
\end{proof}
{\color{black}Theorem~\ref{thm:stronglemma1} shows that real dominance is a property of the
chosen eigenvector, not of the eigenspace. A dominant left eigenvector of a
complex matrix is determined only up to a factor, so even if $w^H$
fails real dominance, the rotated eigenvector $z^H = e^{\mathrm{i}\psi} w^H$
may satisfy it \cite{varga2012}. So, a complex matrix that is not eventually positive may become so
after a suitable rotation. For real matrices, and for complex matrices with real
dominant eigenvectors, the only freedom is a sign, so no such rotation exists.}
 
 The next result provides equivalent conditions to characterize rEEP matrices and clarifies their relationship with the class of PF matrices.
\begin{theorem}\label{thm:stronglemma}
Let $M \in \mathbb{C}^{n \times n}$. Assume $\Re({v})\geq\left | \Im({v}) \right |$ and $\Re({w^H})\geq \left | \Im({w^H}) \right | $, where ${v},~{w^H}$ are the normalized dominant right and left eigenvectors of $M$, respectively. Then, the following statements are equivalent for some scalar $d\geq 0$:
\begin{enumerate}[label=(\roman*)]
\item $ (M+d I) \text{ and } (M+d I)^{H} \in \mathcal{PF}$.
\item $(M+d I)$ is real eventually positive for all $k \geq k_{0}$.
\item $M $ is rEEP.
\end{enumerate}  
\end{theorem}
\begin{proof}  
$(i) \implies (ii)$: The Jordan decomposition of a matrix is given as $M=V J W^H$, where $J$ is the Jordan matrix consisting of all the eigenvalues. Let $\lambda_1$ be the dominant eigenvalue in the square matrix $J$. Then, $M$ is partitioned as:
$$
M=\left[v \mid V_{n, n-1}\right]\left[\begin{array}{l|l}
\lambda_{1} & 0 \\
\hline 0 & J_{n-1}
\end{array}\right]\left[\begin{array}{l}
w^{H} \\
\hline W_{n-1, n}
\end{array}\right],
$$
where $v \text{ and } w$ are the right and left dominant eigenvectors of $M$, normalised such that $w^{H} v=1$. Since  $(M+dI) \text{ and } (M+dI)^{H} \in \mathcal{PF}, \text{ this implies } \Re(v)>0 \text{ and }\Re(w)>0$ from Equation~\eqref{eq:strongPFmatrix}. 
Because ($M+dI\text{ and }(M+dI)^{H} \in \mathcal{PF}$),
\begin{equation}\label{eq: i implies ii}
\frac{(M+dI)^{k}}{\left(\lambda_{1}+d\right)^{k}}=\left[v \mid V_{n-1, n}\right]\left[\begin{array}{c|c}
1 & 0 \\
\hline 0 & \frac{(J'_{n-1})^{k}}{\left(\lambda_{1}+d\right)^{k}}
\end{array}\right]\left[\begin{array}{c}
w^{H} \\
{W_{n-1,n}}
\end{array}\right],
\end{equation}
where $J'_{n-1}=\operatorname{diag}(\lambda_2+d,\ldots,\lambda_n+d)$ 
and $\lambda_{i} \in \operatorname{spec}(M)$.
Thus, by taking the limit $k\to \infty$ on the either sides of Equation~\eqref{eq: i implies ii} and solving yields,
\begin{align*}
\lim_{k \rightarrow \infty} \frac{(M+d I)^{k}}{\left(\lambda_{1}+d\right)^{k}}=v w^{H}.
\end{align*} 
Considering the real part, we have

\begin{equation}\label{eq:Matsteadystate}
\lim _{k \rightarrow \infty} \Re  \left ( \frac{(M+dI)^{k}}{(\lambda_{1}+d)^{k}}\right )=\Re(\alpha {v}{w}^{H}) > \mathbb{0}_{n\times n}.    
\end{equation}
From Equation~\eqref{eq:Matsteadystate} and the definition of rEP matrices, we conclude that $(M+dI)$ is rEP.  



 $(ii)\implies(i):$ Assume that $\Re{((M+dI)^k)}>0$ and define $B=(M+dI)$. The scalar $d\geq 0$ can be chosen such that $B$ is a non-nilpotent matrix \cite{varga2012}. Thus, it follows that $B,B^{H}\in\mathcal{PF}$ which can be proved on the same lines as in Theorem~ $2.3$,\cite{varga2012}. 
 
$(ii)\implies(iii):$ 
We have $\Re\left((M+d I)^{k}\right)> \mathbb{0}_{n \times n}$. Since, 
\begin{align}\label{exp1}
e^{((M+dI)\tau)} &= I + (M+dI)\tau + \frac{(M+dI)^{2} \tau^2}{2!} + \ldots \frac{(M+dI)^{k_{0}} \tau^{k_{0}}}{k_{0}!}\nonumber \\
    &\quad  + h.o.t. ~.
\end{align}
{\color{black} At a large time instant $(\tau \geq \tau_0)$, the higher order terms $(h.o.t.)$ dominate and hence the whole summation becomes real positive as $(M+dI)$ is rEP.
On the other hand, $e^{M\tau}=e^{-d \tau} e^{(M+dI)\tau},$ where $e^{-d \tau} \geq 0$. For all $\tau \geq \tau_0$, the $h.o.t$ in Equation~\eqref{exp1} dominate the sum of the terms until and including $k_0$. Note that $k_0$ is depends on $\tau_0$. 
Hence, $(M+dI)$ is rEEP, as the whole summation $e^{(M+dI) \tau}$ is positive. Also, $M$ inherits the property of rEEP from $(M+dI)$. This follows directly from the fact that  $e^{M \tau}$ can be expressed as $e^{-d \tau}e^{(M+d I)t}$ where $e^{-d \tau}$ is positive. Hence, $M$ is rEEP.}
 
$(iii)\implies(ii):$ Refer to Theorem~3, \cite{saxena2025laplacianflowscomplexvalueddirected}. 
\end{proof}

Theorem~\ref{thm:stronglemma} states that a matrix $M$ is rEEP if the translated matrix $(M+dI)$ follows the strong complex PF property. This analysis directly extends to Laplacian matrices. Consequently, for any Laplacian matrix satisfying the rEEP property, the matrix, $B=dI-L$ satisfies the strong complex PF property.
The example below considers a translated Laplacian matrix and \textcolor{black}{illustrates the possible rotation in dominant eigenvectors, discussed in previous results}: 
\begin{example}\label{ex:rotationofeigenvectors}
Consider 
    \begin{equation*}
     M=\begin{bmatrix}
0.75-1.125\iota  & 0.588-0.147\iota  & 1.073+1.419\iota \\
  -0.55-\iota & 2.176+0.705\iota & 2.147-0.41\iota\\ 
  2.75-0.125\iota & 0.58-0.147\iota & -0.926+0.419\iota \\
\end{bmatrix}.   
    \end{equation*}
\textcolor{black}{For this matrix,} $\lambda_1=3$ is the dominant eigenvalue with a complex-valued dominant right eigenvector. But the dominant left eigenvector $w^H=[0.5-0.43\iota~ 0.65+0.14\iota ~ 0.65+0.14\iota],~\angle(w^H)=[-41.28^{\circ} ~ 12.7^{\circ} ~ 12.4^{\circ}]$ and $\angle(z^{H})=[ 0^{\circ} ~-53.1^{\circ}~-53.1^{\circ}]$. Geometrically, none of these vectors lie in the region $[-45^{\circ},45^{\circ}]$. Thus, the assumption that $\Re(z^{H}) \geq  |\Im(z^{H})|$ is violated. However, assume $\psi = 20^{\circ}$. Then, $\angle(z^{H} e^{\iota \psi})=[20^{\circ} ~  -33.1^{\circ} ~ -33.1^{\circ}]$, and again resorting to the geometric intuition, we note that $z$ follows the property of \emph{real dominance}.
\end{example}

\subsection{Analysis of {Complex-valued Pseudoinverse Laplacians}}
\label{sec:laplacian}
{\color{black} The rEEP property ensures consensus in complex-valued Laplacian flows \cite{saxena2025laplacianflowscomplexvalueddirected}. In contrast, the rEEP property in the case of pseudoinverse Laplacian flows implies consensus solely for undirected and weight-balanced digraphs \cite{aditiECC2025}. Hence, the spectral properties and the steady-state behavior of pseudoinverse Laplacian flows in non-weight-balanced digraphs remain unexplored. We now address this gap.} We examine the eigenspace corresponding to the zero eigenvalue of the Laplacian under the pseudoinverse operation. This analysis subsequently helps us to comment on the convergence properties of pseudoinverse Laplacian flows.
\begin{lemma}\label{lem:invariant null space}
    The set of eigenvectors corresponding to the zero eigenvalue of the Laplacian matrix remains same as of the pseudoinverse Laplacian matrix.
\end{lemma}
\begin{proof}
    For non weight-balanced digraphs, we have $L \mathbb{1}_{n}=\mathbb{0}_{n} \text{ and } w^H L=\mathbb{0}^T_{n}.$ Further, from Def.\ref{def:pseudoinverse},
    \begin{equation}\label{eq:selfadjoint1}
    (L^{\dagger}L)^H=L^{\dagger}L,  
    \end{equation}
    \begin{equation}\label{eq:selfadjoint2}
    (L L^{\dagger})^H=L L^{\dagger},  
    \end{equation}
 Using Equation~\eqref{eq:selfadjoint1},
$ \mathbb{1}^T_{n} L^{\dagger}=  \mathbb{1}^T_{n} L^{\dagger}L L^{\dagger}= w^H (L^{\dagger}L)^H L^{\dagger},$ 
        \begin{align*}
        \mathbb{0}_n^T (L^{\dagger})^H L^{\dagger}=\mathbb{0}^T_n,
    \end{align*}
    Similarly using Equation~\eqref{eq:selfadjoint2}, $L^{\dagger}w= \mathbb{0}_n.$
    \end{proof}
The following result states the relation between the Laplacian and pseudoinverse Laplacian matrices using Lemma \ref{lem:invariant null space}. 
\begin{theorem}\label{Thm:pseudoinverses}
 Consider a strongly connected digraph $\mathcal{G}(A)$ and for $1\leq i,j\leq n$. 
Let $v=\alpha \mathbb{1}_n, \alpha \in \mathbb{C} 
\text{ and } w \in \mathbb{C}^n$ be the right and left eigenvectors associated with the zero eigenvalue of $L$, respectively. Suppose the phase angles, $ \phi_{vi} = \angle (v)_i \text{ and } \phi_{wi} = \angle (w)_i $ lie in $[-45^{\circ},45^{\circ}]$. Then, the following statements are equivalent:
\begin{enumerate}[label=(\roman*)]
\item -$L$ is rEEP.
\item -$L^{\dagger}$ is marginally stable.
\item -$L^{\dagger}$ is rEEP.\\
\end{enumerate}
\end{theorem}
\begin{proof}
$(i) \implies (ii)$: From Lemma \ref{thm:non wt.balanced}, if $-L$ is rEEP, then it is equivalently marginal stable. Hence, $spec(-L)=\left \{ 0, -\lambda _{2},-\lambda _{3},...., -\lambda _{n}\right \}$. Then, we have the spectrum of $L^{\dagger}$ given in Equation~\eqref{eq:spec(Linv)},
\begin{equation}\label{eq:spec(Linv)}
spec(-L^{\dagger})=\left \{ 0, -1/\lambda _{2},-1/\lambda _{3},...., -1/\lambda _{n}\right \},
\end{equation}
where $\lambda _{1},\lambda _{2},...,\lambda _{n}$ are the eigenvalues of $L$ \cite{fontan2023pseudoinverses}. Hence, $L^{\dagger}$ is marginally stable. \\
$(ii) \implies (iii)$: From the previous proof, $-L^{\dagger}$ is marginally stable. Using Jordan canonical form and substituting $\lambda_{1}=0$,\\ $-L^{\dagger}=\begin{bmatrix}
 v^{(1)}|& 
V_{n,n-1}\end{bmatrix} \begin{bmatrix} 0 & 0 \\ 0 & -D^{-1} \end{bmatrix} \begin{bmatrix}
w^{(1)^{H}}\\W_{n-1,n} \end{bmatrix},$

where $-D$ is the diagonal matrix containing eigenvalues (except 0) of $-L$. Then,
\begin{equation}\label{eqn7}
e^{-L^{\dagger}t}=\left[v^{(1)} \mid V_{n, n-1}\right]\left[\begin{array}{c|c}
1 & 0 \\
\hline 0 &  e^{-\lambda_{i}}
\end{array}\right]\left[\begin{array}{c}
w^{(1)^{\mathrm{H}}} \\
\hline W_{n-1, n}
\end{array}\right]
\end{equation} 
Equation~\eqref{eqn7} holds for all the cases of degenerate and non-degenerate eigenvalues as the exponential decay is faster than the growth of the corresponding polynomial. 
 In case of $-L^{\dagger}$, the eigenvectors corresponding to zero eigenvalue remain same. Hence, $-L^{\dagger}$ is rEEP.\\
$(iii) \implies (ii)$: This can be proved on the same lines as in the proof of Theorem~3 of \cite{saxena2025laplacianflowscomplexvalueddirected}.
\end{proof}\vspace{-2mm}
An example is provided to validate the previous result. 
\begin{example} 
Consider $L \text{ and }L^{\dagger}$, whose real parts are shown in Figure~\ref{fig:heatmapsL} via heatmaps.
\begin{figure}[H]
    \centering
    \begin{subfigure}{0.2\textwidth}
    \centering
        \includegraphics[scale=0.612]{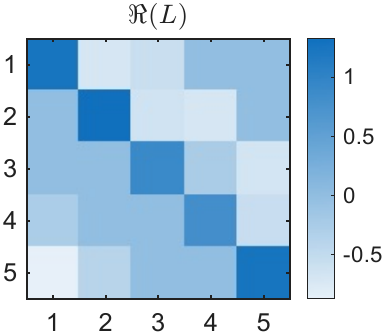}
        \caption{$\Re(L)$}
        \label{fig:Linv_realcase10}
        \end{subfigure}
    \hspace{0.34cm} 
    \begin{subfigure}{0.2\textwidth}
    \centering
\includegraphics[scale=0.612]{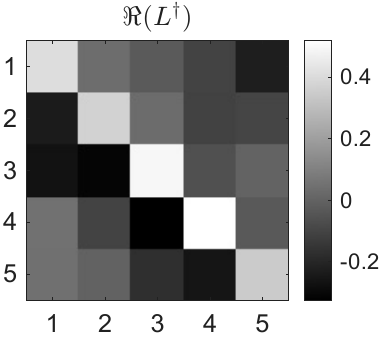}
        \caption{$\Im(L^{\dagger})$}
        \label{fig:Linv_imagcase10}
          \end{subfigure}     
    \caption{\small Heatmaps showing the matrix structure of $L \text{ and } L^\dagger$ for real and imaginary parts respectively}
    \label{fig:heatmapsL}
\end{figure}\vspace{-4mm}
\begin{figure}[H]
    \centering
    \begin{subfigure}{0.2\textwidth}
    \centering
        \includegraphics[scale=0.612]{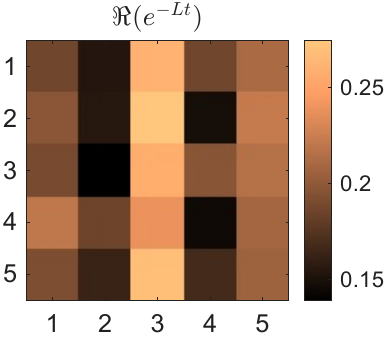}
        \raggedright
        \caption{$\Re(e^{-Lt})$}
        \label{fig:L_realeep case10}
        \end{subfigure}
    \hspace{0.56cm} 
    \begin{subfigure}{0.2\textwidth}
    \centering
\includegraphics[scale=0.612]{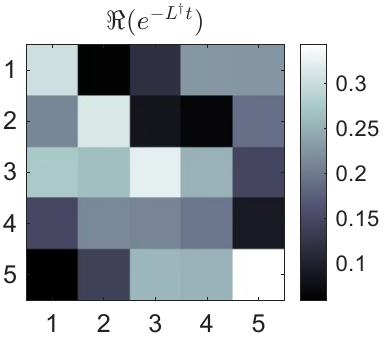}
\raggedright
        \caption{$\Re(e^{-L^{\dagger}t}$)}
        \label{fig:reep heatmap}
          \end{subfigure}     
    \caption{\small The matrix structure of $\Re(e^{-Lt})$ and $\Re(e^{-L^{\dagger}t}$) at $t=3$ and $t=4$ respectively. Values on the colorbar represent the entire range that encompasses all the values of a matrix, and all of these values are positive implying rEEP property.}
    \label{figs:case10 realEEP}
\end{figure}
\end{example} 
The following remarks are in order for the Laplacians, pseudoinverse Laplacians, and real parts of exponential maps shown in Figures~\ref{fig:L_realeep case10} and \ref{fig:reep heatmap}, respectively. First, notice that the real parts vary from positive to negative values for the Laplacian as well as the pseudoinverse Laplacian. Since the spectrum of the Laplacian matrix has a simple zero eigenvalue, the negated Laplacian matrix as well as the negated pseudoinverse Laplacian is rEEP.
Thus, it is evident from Figure~\ref{figs:case10 realEEP} that the rEEP property is preserved under the operation of pseudoinverse for Laplacian matrices, as proved in Theorem~\ref{Thm:pseudoinverses}.
In contrast to the results in \cite{aditiECC2025}, the pseudoinverse Laplacian flows do not converge in the case of non-weight-balanced digraphs. It happens so because from Theorem~\ref{Thm:pseudoinverses}, we conclude that the property of rEEP is conserved under the pseudoinverse operation. But referring $x(t) =e^{-L^{\dagger }t} x(0) $ as the solution of pseudoinverse Laplacian flow given by $\dot{x}(t)=-L^{\dagger }x(t):$
\begin{equation*}
\lim _{t \rightarrow \infty} e^{-L^{\dagger }t} = w \underbrace{v^{H} \cdot x_{0}}_{\text {constant=k }} 
\end{equation*}
\begin{equation*}
 \lim _{t \rightarrow \infty}   x(t)=kw \implies ~x(t) \notin \operatorname{span}\left\{\mathbb{1} _{n}\right\} \text{because } w\notin \operatorname{span}\left\{\mathbb{1} _{n}\right\}.
\end{equation*}
Next, we demonstrate that the eigenspace, along with the zero eigenvalue remains same when we rotate all the elements of the Laplacian matrix with some angle of rotation, $\psi$. In particular, a global phase rotation of the Laplacian matrix rotates its non-zero spectrum but leaves the zero eigenvalue and the eigenspace completely invariant.
\begin{corollary}\label{Proposition:invariant null space}
The set of eigenvectors corresponding to the zero eigenvalue of remains invariant under rotation of the Laplacian matrix; i.e., if all the elements of the matrix are rotated by a rotation angle of $\psi$.
\end{corollary}
\begin{proof}
Let $L\in\mathbb{C}^{n\times n}$ be a complex-valued Laplacian matrix. And $L_{\psi}=e^{\iota \psi}L$ be the rotated Laplacian with an angle $\psi\in \mathbb{C}$ and $t \in \mathbb{R}$. Let $v \in \mathbb{C}^n$ such that 
$Lv=0$. It then follows that
\begin{equation}
L_{\psi}v=e^{\iota \psi}Lv=0.    
\end{equation}
Hence $v$ is the eigenvector corresponding to the zero eigevanlue of $L_{\psi}$
Conversely, because $e^{\iota \psi}\neq 0$, $L_{\psi}v=0$ implies $Lv=0$. Thus ${\ker(L_{\psi})=\ker(L)}$. In other words, eigenspaces corresponding to the rotated and actual Laplacians coincide. 
Similar argument holds for left eigenspaces corresponding to the zero eigenvalue as well. 
\end{proof}

Corollary~\ref{Proposition:invariant null space} implies that the rEEP property is also preserved under a uniform rotation. This is because the rEEP property depends only on the eigenvectors corresponding to the zero eigenvalue of the complex-valued Laplacian.   
\begin{remark}
The property of rEEP is not always invariant under the global phase rotation. This is because of the change in the non-zero eigenvalues in the spectrum. While the zero eigenvalue and its eigenspace remain unchanged, the non-zero eigenvalues are rotated by the angle, $\psi$, which can shift them in the left half of the s-plane. Thus, the rEEP property is preserved only if the non-zero eigenvalues of the negated Laplacian remain in the right half of the complex plane after rotation. Hence, the property of rEEP depends on the full spectrum and does not guarantee consensus in all cases. 
\end{remark}
\section{CONCLUSIONS AND FUTURE WORK }
\textcolor{black}{We extend real eventual exponential positivity (rEEP) to pseudoinverses of complex-valued Laplacians on digraphs. We show that rEEP is preserved under the pseudoinverse, and that it is determined entirely by the relative phases of the edge-weights. We also show that the null space of the Laplacian is invariant under a uniform rotation of the edge-weights. Altogether, these results provide a foundation for further analysis of consensus and stability in complex-valued networked systems. Our future studies will include deriving the conditions to achieve rEEP Laplacians, without imposing the real dominance on the eigenvectors corresponding to the zero eigenvalue of complex-valued Laplacians.}
       
\bibliography{References}

@INPROCEEDINGS{aditiECC2025,
  author={Saxena, Aditi and Tripathy, Twinkle and Anguluri, Rajasekhar},
  booktitle={2025 European Control Conference (ECC)}, 
  title={Are the flows of complex-valued Laplacians and their pseudoinverses related?}, 
  year={2025},
  volume={},
  number={},
  pages={653-658},
  doi={10.23919/ECC65951.2025.11186905}}

@INPROCEEDINGS{aditiICC2024,
  author={Saxena, Aditi and Tripathy, Twinkle and Anguluri, Rajasekhar},
  booktitle={2024 Tenth Indian Control Conference (ICC)}, 
  title={Real Eventual Exponential Positivity of Complex-valued Laplacians: Applications to Consensus in Multi-Agent Systems}, 
  year={2024},
  volume={},
  number={},
  pages={226-231},
  doi={10.1109/ICC64753.2024.10883735}}

@article{DONG20161,
title = {{L}aplacian matrices of general complex weighted directed graphs},
journal = {Linear Algebra and its Applications},
volume = {510},
pages = {1-9},
year = {2016},
issn = {0024-3795},
doi = {https://doi.org/10.1016/j.laa.2016.08.011},
url = {https://www.sciencedirect.com/science/article/pii/S002437951630338X},
author = {Jiu-Gang Dong and Lin Lin},
}

@article{tian2024structuralbalance,
author = {Tian, Yu and Lambiotte, Renaud},
title = {Structural Balance and Random Walks on Complex Networks with Complex Weights},
journal = {SIAM Journal on Mathematics of Data Science},
volume = {6},
number = {2},
pages = {372-399},
year = {2024},
doi = {10.1137/23M1584265},

URL = { 
    
        https://doi.org/10.1137/23M1584265
},
eprint = { 
    
        https://doi.org/10.1137/23M1584265
}
}

@article{dong2014complex,
  title={Complex {L}aplacians and applications in multi-agent systems},
  author={Dong, Jiu-Gang and Qiu, Li},
  journal={arXiv preprint arXiv:1406.1862},
  year={2014}
}

@article{dorfler2012synchronization,
  title={Synchronization and transient stability in power networks and nonuniform {K}uramoto oscillators},
  author={Dorfler, Florian and Bullo, Francesco},
  journal={SIAM Journal on Control and Optimization},
  volume={50},
  number={3},
  pages={1616--1642},
  year={2012},
  publisher={SIAM}
}

@article{olfati2007consensus,
  title={Consensus and cooperation in networked multi-agent systems},
  author={Olfati-Saber, Reza and Fax, J Alex and Murray, Richard M},
  journal={Proceedings of the IEEE},
  volume={95},
  number={1},
  pages={215--233},
  year={2007},
  publisher={IEEE}
}

@article{varga2012,
  title={On the {P}erron--{F}robenius theory for complex matrices},
  author={Noutsos, Dimitrios and Varga, Richard S},
  journal={Linear algebra and its applications},
  volume={437},
  number={4},
  pages={1071--1088},
  year={2012},
  publisher={Elsevier}
}

@article{tian2026generalizing,
  title={Generalizing Perron--Frobenius theory and eigenvector-based centralities to networks with complex edge weights},
  author={Tian, Yu and Porter, Mason A and B{\"o}ttcher, Lucas},
  journal={arXiv preprint arXiv:2606.12026},
  year={2026}
}

@article{2012moore,
  title={The {M}oore--{P}enrose pseudoinverse: A tutorial review of the theory},
  author={Barata, Jo{\~a}o and Carlos Alves and Hussein, Mahir Saleh},
  journal={Brazilian Journal of Physics},
  volume={42},
  pages={146--165},
  year={2012},
  publisher={Springer}
}

@book{bullo2018lectures,
  title={Lectures on network systems},
  author={Bullo, Francesco},
  volume={1},
  year={2018},
  publisher={CreateSpace}
}

@article{bottcher2024complex,
  title={Complex networks with complex weights},
  author={B{\"o}ttcher, Lucas and Porter, Mason A},
  journal={Physical Review E},
  volume={109},
  number={2},
  pages={024314},
  year={2024},
  publisher={APS}
}

@inproceedings{fontan2021properties,
  title={On the properties of {L}aplacian pseudoinverses},
  author={Fontan, Angela and Altafini, Claudio},
  booktitle={2021 60th IEEE Conference on Decision and Control (CDC)},
  pages={5538--5543},
  year={2021},
  organization={IEEE}
}

@article{fontan2022multiagent,
  title={Multiagent consensus over time-invariant and time-varying signed digraphs via eventual positivity},
  author={Fontan, Angela and Wang, Lingfei and Hong, Yiguang and Shi, Guodong and Altafini, Claudio},
  journal={IEEE Transactions on Automatic Control},
  volume={68},
  number={9},
  pages={5429--5444},
  year={2022},
  publisher={IEEE}
}

@article{lin2013leader,
  title={Leader--follower formation via complex {L}aplacian},
  author={Lin, Zhiyun and Ding, Wei and Yan, Gangfeng and Yu, Changbin and Giua, Alessandro},
  journal={Automatica},
  volume={49},
  number={6},
  pages={1900--1906},
  year={2013},
  publisher={Elsevier}
}

@article{saxena2026emotionsspreadaffectdecisionmaking,
  title={How Do Emotions Spread And Affect Decision-Making In Social Networks? An Opinion Dynamics Perspective},
  author={Saxena, Aditi and Tripathy, Twinkle},
  journal={arXiv preprint arXiv:2610.22334},
  year={2026}
}

@article{saxena2025laplacianflowscomplexvalueddirected,
  title={Laplacian Flows in Complex-valued Directed Networks: Analysis, Design, and Consensus},
  author={Saxena, Aditi and Tripathy, Twinkle and Anguluri, Rajasekhar},
  journal={arXiv preprint arXiv:2509.04196},
  year={2025}
}

@article{fontan2023pseudoinverses,
  title={Pseudoinverses of signed {L}aplacian matrices},
  author={Fontan, Angela and Altafini, Claudio},
  journal={SIAM Journal on Matrix Analysis and Applications},
  volume={44},
  number={2},
  pages={622--647},
  year={2023},
  publisher={SIAM}
}

@inproceedings{altafini2019investigating,
  title={Investigating stability of {L}aplacians on signed digraphs via eventual positivity},
  author={Altafini, Claudio},
  booktitle={2019 IEEE 58th Conference on Decision and Control (CDC)},
  pages={5044--5049},
  year={2019},
  organization={IEEE}
}
\bibliographystyle{unsrt}
\end{document}